\documentclass[letter]{aa} 
\usepackage{graphicx}
\usepackage{txfonts}
\begin{document}

   \title{The width of Herschel filaments varies with distance}

   \author{G. V. Panopoulou,
          \inst{1}\fnmsep\thanks{Hubble fellow}
          S. E. Clark,\inst{2,3}
A. Hacar,\inst{4}
F. Heitsch,\inst{5}
J. Kainulainen,\inst{6}
E. Ntormousi,\inst{7}
D. Seifried,\inst{8}
\and R. J. Smith\inst{9}
          }

   \institute{California Institute of Technology, Mail Code 350-17, 1200 E. California Blvd, Pasadena, CA 91125, USA\\
              \email{panopg@caltech.edu}
         \and
             Department of Physics, Stanford University, Stanford, California 94305, USA
             \and
             Kavli Institute for Particle Astrophysics \& Cosmology, P. O. Box 2450, Stanford University, Stanford, CA 94305, USA
             \and
             Department of Astrophysics, University of Vienna, T{\"u}rkenschanzstrasse 17, 1180, Vienna, Austria
             \and
             Department of Physics \& Astronomy, University of North Carolina - Chapel Hill, Chapel Hill, North Carolina 27599, USA
             \and
             Chalmers University of Technology, Department of Space, Earth and Environment, SE-412 93, Gothenburg, Sweden 
             \and
             Scuola Normale Superiore di Pisa, Piazza dei Cavalieri, 7, 56126 Pisa, Italy
             \and
             University of Cologne, I. Physical Institute, Z\"ulpicher Str. 77, 50937 Cologne, Germany
             \and
             Jodrell Bank Centre for Astrophysics, Department of Physics and Astronomy, University of Manchester, Oxford Road, Manchester M13 9PL, UK
             }


 
  \abstract
   {Filamentary structures in nearby molecular clouds have been found to exhibit a characteristic width of 0.1 pc, as observed in dust emission. Understanding the origin of this universal width has become a topic of central importance in the study of  molecular cloud structure and the early stages of star formation.}
   {We investigate how the recovered widths of filaments depend on the distance from the observer by using previously published results from the \textit{Herschel} Gould Belt Survey.}
   {We obtained updated estimates on the distances to nearby molecular clouds observed with \textit{Herschel} by using recent results based on 3D dust extinction mapping and \textit{Gaia}. We examined the widths of filaments from individual clouds separately, as opposed to treating them as a single population. We used these per-cloud filament widths to search for signs of variation amongst the clouds of the previously published study. }
   {We find a significant dependence of the mean per-cloud filament width with distance. The distribution of mean filament widths for nearby clouds is incompatible with that of farther away clouds. The mean per-cloud widths scale with distance approximately as 4--5 times the beam size. We examine the effects of resolution by performing a convergence study of a filament profile in the \textit{Herschel} image of the Taurus Molecular Cloud. We find that resolution can severely affect the shapes of radial profiles over the observed range of distances.}
   {We conclude that the data are inconsistent with 0.1 pc being the universal characteristic width of filaments.}

   \titlerunning{Herschel filament widths vary with distance}
   \authorrunning{Panopoulou et al.}
   
   \maketitle
   
%

\section{Introduction} \label{sec:intro}

The formation of stars appears to occur preferentially within filamentary structures \citep{Andre2014,Konyves2015}. Special attention has been given to one morphological property of these structures: their width. In models of idealized hydrostatic cylinders, the radius is related to the stability of the structure \citep{Ostriker1964}. In particular, the radius of the column density profile is expected to scale inversely with the column density following the thermal Jeans length. \citet{Arzoumanian2011} analyze a large sample of filaments in \textit{Herschel} dust continuum images and show that the observed widths of filaments are almost independent of the column density and are uncorrelated with the Jeans length -- contrary to theoretical expectations. 

Despite the wide range of filament column densities, filament widths are found to follow a narrow distribution, which  peaks at $\sim$~0.1~pc with a spread of only a factor of 2 \citep{Arzoumanian2011}.
This surprising finding has led to the proposition that filaments show a characteristic width -- one that is universal among clouds with drastically different properties \citep[e.g., star formation rate, mean column density,][]{Arzoumanian2011,Andre2014}. Recently, \citet{Arzoumanian2019} extended the analysis of filament widths to a much larger sample of filaments in the \textit{Herschel} Gould Belt Survey (HGBS), finding results in agreement with their earlier study.
Multiple theoretical models have been proposed to explain the observed distribution of widths and the apparent independence with the column density \citep{Fischera2012a,Fischera2012b,Hennebelle2013,HennebelleAndre2013,Federrath2016,Auddy2016,Federrath2021,Priestley2021}. To date, no model has been able to reproduce the properties of the distribution over the wide range of filament column densities in the sample of \citet{Arzoumanian2019}.

The presence of a characteristic width has been called into question from several investigations. \citet{Panopoulou2017} show that commonly adopted choices in the analysis of filament radial profiles lead to significant biases in the resulting width distribution. First, the width of \textit{Herschel} filaments was originally determined from the full-width-at-half-maximum (FWHM) of a single radial profile: one that results from averaging the contribution of equidistant points at each radius along the filament spine. The reported distribution of widths is thus a distribution of the mean filament widths, leading to an artificially narrow spread as a result of the central limit theorem. When considering widths measured at all points along a filament's crest, broader distributions are invariably found \citep[with a spread 2-3 times that of the crest-averaged distribution,][]{Panopoulou2017,Arzoumanian2019,Suri2019}. Second, the determination of the filament FWHM has a strong dependence on the choice  of the maximum radial distance within which the fit is performed \citep{Smith2014}.
\citet{Ossenkopf-Okada2019} perform an  independent analysis of \textit{Herschel} data using a wavelet decomposition and do not find signs of a characteristic width common to all of the clouds in their study.
Recently, \citet{Louvet2021} investigated the effect of the telescope beam size on the core mass function (CMF), finding that both the peak of the CMF and the radial extent of filaments are dependent on the resolution.

\citet{Juvela2012} study how telescope resolution can affect the properties of recovered filament profiles by employing magneto-hydrodynamical (MHD) simulations and radiative transfer post-processing. The resulting filament widths are mildly affected (10\% level) unless the structures are placed at distances $\gtrsim$400 pc (beyond which the structures become unresolved), they are affected by background confusion, or they have complex dust opacity (in which case biases of $\sim$40\% are found).
Observations treating the effect of varying dust optical properties have also found slight variations in filament widths compared to the case of the assumed simple opacity on the order of 60\% \citep{Howard2019}. The level of bias caused by resolution on filament widths in the aforementioned works is model-dependent; for example, the 0.01-pc-wide filaments simulated
by \citet []{Seifried2017} have observed widths a few times wider than their true value.

In this Letter, we revisit the original data that support the presence of a characteristic width of 0.1 pc. Using the most recent developments in the determination of molecular cloud distances based on \textit{Gaia}, we revised the estimates of filament widths published from the HGBS survey.
We demonstrate that the mean filament width increases as a function of distance (Sect. \ref{sec:widths_with_distance}). This trend refers to the ensemble average of widths over the population of filaments in a cloud. We investigate whether the trend could be related to telescope resolution though a convergence study of a single filament profile (Sect. \ref{sec:Plummer}). We discuss our findings and conclude in Sect. \ref{sec:conclusions}.

\section{Data}
\label{sec:data}

\subsection{Literature measurements of filament widths}

\citet{Arzoumanian2019} identified filaments on column density maps from \textit{Herschel} for eight clouds in the HGBS. We briefly summarize the salient points in their analysis leading to the determination of filament widths. 

In each image, the crest of filamentary structures (skeleton) was obtained by using the DisPerSe algorithm \citep{Sousbie2011}. For each filament, a single radial column density profile was created by taking the median of all points that are equidistant from the crest along the length of the filament (we refer to this as the crest-averaged profile). At some distance from the crest, the profile flattens and merges with the background. The radius at which this happens is denoted as $r_{\rm out}$. The width of the filament profile within $r_{\rm out}$ of the crest was measured in two ways. First, the authors found the radius where the profile drops to half-maximum of the crest-averaged profile after background subtraction (half-radius, $hr$, in their notation). The width is defined as the half-diameter $hd = 2hr$. Second, a Gaussian function plus background was fit within 1.5 $hr$, and the width of the filament is the resulting FWHM of the Gaussian.
For each cloud, the distribution of filament widths was constructed. \citet{Arzoumanian2019} calculated a "deconvolved" width, or half-diameter, as follows: $hd_{dec} = \sqrt{hd^2 - HPBW^2}$ (HPBW is the telescope half-power-beam-width).

In this paper we use the nonparametric estimation of filament width, reported as the "deconvolved" half-diameter,  $hd_{dec}$ in Table 3 of \citet{Arzoumanian2019}. In our notation, we define $hd_{dec}$ to be $\rm FWHM_{dec}$. When necessary, we convert the "deconvolved" width, $\rm FWHM_{dec}$, to the observed width, $\rm FWHM_{obs}$, following \citet{Arzoumanian2019}:
\begin{equation}
\rm FWHM_{obs} = \rm \sqrt{\rm FWHM_{dec}^2 + HPBW^2},
\label{eqn:deconv}
\end{equation}
where HPBW, the telescope half-power-beam-width, is equal to 18.2\arcsec. We stress that calculating $\rm FWHM_{dec}$ as in \citet{Arzoumanian2019} does not accurately correct for the convolution with the beam, as we show in Appendix \ref{sec:appendixTau}, but we chose to use $\rm FWHM_{dec}$ to facilitate comparison with their work.
In addition to the $hd_{dec}$, we also used  the values of $2 r_{\rm out}$ as well as the spread of the per-cloud distribution of $hd_{dec}$ --which we denote as $\sigma({\rm FWHM_{dec}}$)-- as provided in their table 3. 


\subsection{Cloud distances}

We used the latest 3D dust extinction maps based on \textit{Gaia} for the determination of distances to clouds in the \citet{Arzoumanian2019} sample. \citet{Zucker2020} have provided highly accurate distance measurements (to within $\sim 5\%$) for a subset of the clouds in this sample, namely: IC\,5146, Orion\,B, Taurus L1495/B213, and Ophiuchus. While they also provide estimates for the Aquila Rift, the Pipe Nebula, and the Polaris Flare, these are based on sightlines passing outside the area covered by \textit{Herschel}. We therefore reanalyzed data from 3D dust extinction toward these three clouds, as well as Musca, which does not have a recent distance estimate in the literature, to determine the distance to the filamentary structures seen in the \textit{Herschel} images. For this, we used the \citet{Leike2020} 3D dust map which provides the highest distance resolution among the existing maps within the Solar neighborhood, as described in Appendix \ref{sec:appendix}. Distances to the Polaris Flare and IC 5146 are the most discrepant between the updated  measurements and the default values adopted in \citet{Arzoumanian2019}. The results are summarized in Table \ref{tab:clouddist}. Throughout the text, updated distance estimates are denoted as $d_{\rm new}$, while those used in \citet{Arzoumanian2019} are denoted as $d_{\rm old}$.

\begin{figure}
    \centering
    \includegraphics[scale=1]{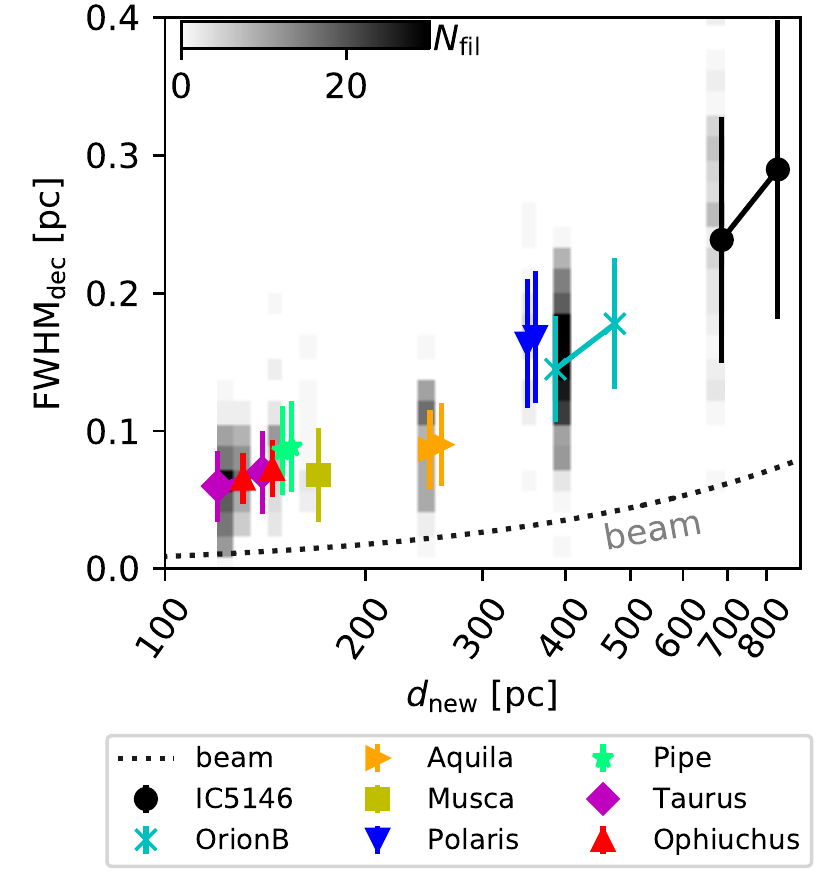}
    \caption{Mean "deconvolved" filament width for each cloud analyzed by \citet{Arzoumanian2019}, as a function of the updated cloud distance ($d_{\rm new}$). A colored symbol shows the ensemble average width of filaments within a cloud. The vertical lines span the $\pm1\sigma$ range of the distribution of the filament widths of that cloud. Two points mark the full range of distances to a cloud, when such measurements are available. A dotted line shows the \textit{Herschel} beam size. The background shows a distribution of the filament widths for each cloud, drawn from a Gaussian distribution with the number of measurements per cloud reported in the original study (see text). We only show a realization for the nearest distance limit of each cloud for clarity. 
    }
    \label{fig:wa_dist}
\end{figure}

\section{Dependence of filament widths on  distance}
\label{sec:widths_with_distance}

Using the new cloud distances, $d_{\rm new}$, we rescaled the per-cloud mean filament widths from \citet{{Arzoumanian2019}} (from their table 3, see Section \ref{sec:data}) to obtain revised estimates of FWHM$_{\rm dec}$ for each cloud. The same operation was performed to the spread of the distribution of widths for each cloud. Fig. \ref{fig:wa_dist} shows the mean "deconvolved" filament width, FWHM$_{\rm dec}$, as a function of $d_{\rm new}$, for each cloud in the \citet{Arzoumanian2019} sample separately. Each data point represents an ensemble average over the set of filaments identified in a cloud by their study, and error-bars denote the standard deviation of the distribution of filament widths in each cloud. The mean filament width systematically increases by a factor of $\sim 4$ when comparing the nearest and farthest clouds. For each cloud we use the full range of likely distances, as obtained in Appendix \ref{sec:appendix}, to demonstrate the full error budget. 
We find a similar trend in the original data as published in \citet{Arzoumanian2019} (which show a factor of $\sim 2$ increase). For example, using the values reported in table 3 of \citet{Arzoumanian2019}, filament widths in IC 5146 (with a mean of 0.13 pc and $\sigma$ of 0.04 pc) are more than one standard deviation larger than those in Taurus (with a mean of 0.06 pc and a spread of 0.02 pc). We have verified that the trend is also seen in the widths obtained by Gaussian fits provided in table 3 of \citet{Arzoumanian2019}.

Clouds at distances larger than 300 pc (Polaris, Orion, IC5146) clearly show higher values for the mean width. If one were to combine filament widths from all clouds, as in \citet{Arzoumanian2019}, the resulting distribution would have a mean of $\sim$ 0.1 pc. This is noted in their section 4.1 when alternative distances are considered. However, it is clear from Fig. \ref{fig:wa_dist} that the clouds that are farther than $\sim$300~pc pull the mean of the distribution toward higher values. In particular, the 234 filaments of Orion B constitute 40\% of the total number of filaments in the sample and  significantly contribute to the larger values around 0.1 pc. 

To demonstrate the level of discrepancy, we compared the distributions of the two clouds with the largest number of filaments (Orion B and Taurus, with 234 and 110 filaments, respectively) via a two-sided Kolmogorov-Smirnov (K-S) test. We used the reported mean width, the $\pm 1 \sigma$ standard deviation, and the number of identified filaments, $N\rm _{fil}$, to draw mock filament widths from a normal distribution. Fig. \ref{fig:wa_dist} shows one random realization of filament widths drawn as described above for all clouds (background grayscale), using the same number of filaments per cloud as the original paper. The K-S test rejects the hypothesis that the distribution of widths from filaments in Orion is drawn from the same distribution as that of Taurus, with a p-value $< 10^{-40}$. This is true for both the original widths as well as the rescaled values after updating cloud distances. 

Finally, we also examined the spread of the per-cloud distribution of filament widths. If all clouds had the same intrinsic spread of filament widths, we would expect to find the reported $\sigma$ of the distribution of FWHM$_{\rm dec}$ $\sigma$(FWHM$_{\rm dec}$) to be independent of distance\footnote{This expectation holds if the intrinsic spread is measurable with \textit{Herschel}, e.g., not limited by resolution.}.
Contrary to that expectation, the per-cloud spread $\sigma$(FWHM$_{\rm dec}$) are correlated with the cloud distance, and they are found to increase from 0.026 for the nearest cloud to 0.09 pc for the farthest cloud (Pearson correlation coefficient of 0.94 and a p-value of 4 $\times 10^{-4}$). It is unlikely that the different sample sizes are driving this scaling, as the $\sigma$(FWHM$_{\rm dec}$) are uncorrelated with the number of filaments per cloud, $N_{\rm fil}$.

In summary, we have found that the mean and the spread of the per-cloud distribution of filament widths significantly depend on distance from the observer. A K-S test rejects the null hypothesis that the distribution of widths in nearby clouds is consistent with that of farther away clouds. These findings contradict the interpretation of the mean width of 0.1 pc, averaged over all clouds, as being representative of the whole filament sample (i.e., universal).

\section{Resolution may strongly affect filament profiles}
\label{sec:Plummer}

The apparent increase of the mean filament widths as a function of distance (Fig. \ref{fig:wa_dist}) suggests that filament profiles are not resolved. Yet, filament FWHM$\rm _{dec}$ are several times larger than the beam size of 18.2\arcsec (as can be seen by comparing the data with the line of Fig. \ref{fig:wa_dist}). 
To understand this apparent contradiction, we investigated the effect of the resolution on the profile of a filament in the nearby Taurus molecular cloud. We tested the following hypothesis: Is resolution, in principle, capable of producing as significant a rise of filament width with distance as observed? To answer this question, we performed a simple experiment: we progressively reduced the resolution of the map of the Taurus main filament, effectively "observing" it with angular resolution corresponding to the original \textit{Herschel} beam at larger distances \citep[i.e., we performed a convergence test as in the CMF study of][]{Louvet2021}. We then measured the FWHM of the filament at these different resolutions.

As we are interested in understanding the effect of the beam size, we used the observed FWHM$\rm _{obs}$, not the "deconvolved" $\rm FWHM_{dec}$ (Eq. \ref{eqn:deconv}), as well as the updated distance estimates, $d_{\rm new}$ (Table \ref{tab:clouddist}).
The measurement of FWHM$_{\rm obs}$ involves two main operations: (a) convolution with the telescope beam and (b) truncation of the profile at radii larger than $r_{\rm out}$ (Section \ref{sec:data}). First, we numerically convolved the column density image of Taurus to achieve resolutions equivalent to 0.023 pc, 0.037 pc, and 0.067 pc (corresponding to a beam size of 18.2\arcsec at distances of 260 pc, 423 pc, and 762 pc). Using the same filament skeleton for all images, we constructed the filament's radial profile, determined $r_{\rm out}$, and measured FWHM$\rm _{obs}$. We refer the reader to Appendix \ref{sec:appendixTau} for more details.

\begin{figure}
    \centering
    \includegraphics[scale=0.96]{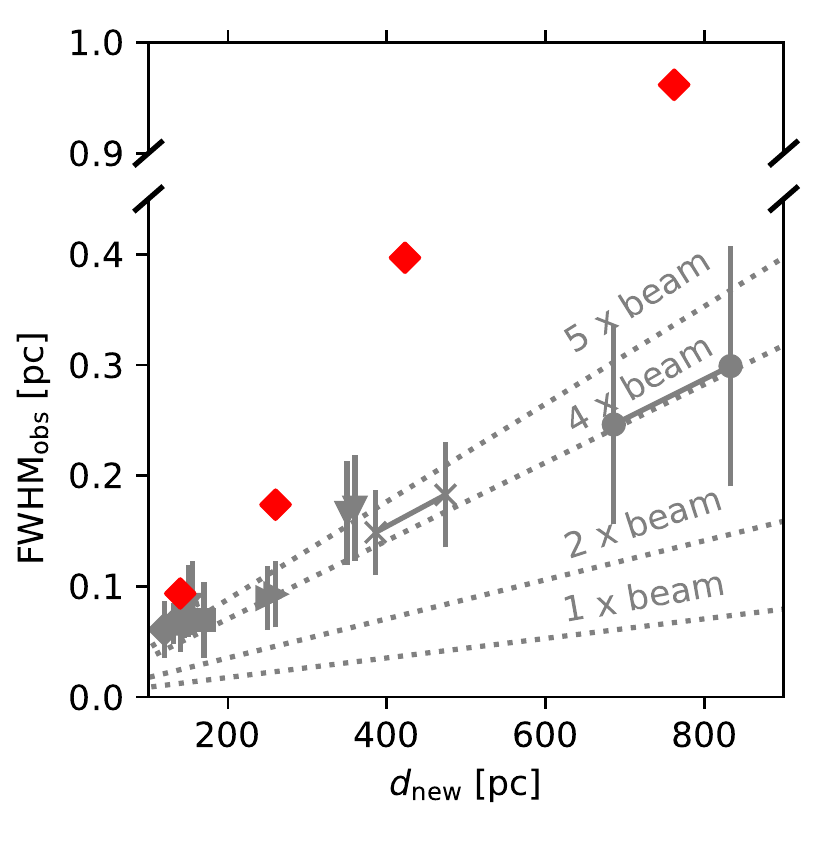}
    \caption{Comparison between the observed dependence of FWHM$_{\rm obs}$ on the distance (gray data points, symbols as in Fig. \ref{fig:wa_dist}) and that of a radial profile of a filament in Taurus, after reducing the angular resolution of the \textit{Herschel} map to correspond to the physical resolution of a 18.2\arcsec\, beam at the observed cloud distances (red diamonds). Dotted lines mark 1, 2, 4, and 5 times the beam size.}

    \label{fig:tau_dist}
\end{figure}

In Fig. \ref{fig:tau_dist} we compare the $\rm FWHM_{obs}$ obtained for this resolution study with the per-cloud mean $\rm FWHM_{obs}$, as a function of cloud distance. 
The effect of resolution dramatically changes the $\rm FWHM_{obs}$ of the Taurus filament profile (red diamonds), with values increasing by a factor of 10 from the original image (140 pc) to the largest distance of 762 pc. While the per-cloud mean $\rm FWHM_{obs}$ do rise with distance, this rise is shallower than that of the single profile of the resolution study. We note that the reduction of angular resolution is not necessarily equivalent to placing the same cloud at different distances. However, this convergence study allows us to examine whether all filaments observed with the same resolution in physical units have the same width of 0.1 pc. This does not seem to be the case: if the Taurus filament had been observed with a physical resolution of 0.067 pc (as filaments in IC 5146) instead of 0.012 pc, it would appear to have a width of 0.99 pc compared to the mean width of 0.24 pc of filaments in IC 5146.

We have thus shown that resolution can produce an increase in the FWHM$\rm _{obs}$ of a single filament by a factor of $\sim 10$, which is much larger than the increase in the per-cloud mean FWHM$\rm _{obs}$ (factor of 4). The observed rise in the per-cloud mean FWHM$\rm _{obs}$ with distance could, at least partially, be explained by resolution. The remarkably linear dependence of FWHM$\rm _{obs}$ on the distance of $\sim 4 \times \rm beam$ at large distances would suggest that indeed filaments are not resolved.
Resolution and beam confusion can lead to the detection of structures of a few times the beam size; for example, cloud diameters were found to
be 3 times the beam size in \citet[][]{Verschuur93} as a result of the hierarchical nature
of the medium. Further work is needed to reliably disentangle possible effects of resolution from a possible intrinsic variation of filament widths. The mean width of the entire cloud sample of 0.1 pc may not be robust to changes in resolution.
Verifying or disproving this, however, is beyond the scope of the present work.

\section{Conclusions}
\label{sec:conclusions}

We have revisited the observational results of \citet{Arzoumanian2019} on filament widths in the HGBS. By examining the data of different clouds separately, we have unveiled a trend that has been hidden in the data: a dependence of the per-cloud mean filament widths on the distance from the observer. While the ensemble average widths over all clouds is 0.1 pc, this is not a representative
statistic. The distribution of mean filament widths for nearby clouds is incompatible with that of farther away clouds. In addition, the spread of per-cloud filament widths depends on distance. The data thus contradict a universal width of 0.1 pc with a spread of $\sim$ 2 for all clouds, as originally inferred by \citet{Arzoumanian2011}. 

The scaling of FWHM with distance is reminiscent of that found in other \textit{Herschel} surveys \citep{Schisano2014,RiveraIngraham2016}, recognised as the effect of resolution.
Even though the observed filament widths are multiple times the beam size, we have demonstrated that the effect of resolution on the shape of an observed filament profile is not negligible. 
We considered the case of the profile of a filament in the Taurus molecular cloud. We performed a resolution study by convolving the column density image of Taurus with varying beam sizes. The resolution has a dramatic effect on the recovered filament width: the width increases with distance more steeply than the per-cloud mean widths. In combination with the almost linear scaling of mean filament widths with distance (4-5 times the beam size), our results strongly suggest that resolution biases the measurement of filament widths.

Our results are in agreement with a growing body of evidence showing that filament radial profiles exhibit more complexity than can be captured by the picture of a "characteristic" width. Filaments show substructure when observed with spectral line tracers \citep{Hacar2013} and can appear significantly narrower when observed at a higher resolution than \textit{Herschel} \citep[e.g., with interferometers targeting spectral lines of  dense gas tracers,][]{Fernandez-Lopez2014,Hacar2018,Monsch2018}. Filament radial profiles can vary by factors of 2-10 from one end of the structure to the other \citep{Juvela2012coldcores,Suri2019}, and averaging is shown to bias the shape of the resulting profile \citep{Whitowrth2021}.

By understanding the inherent biases in characterizing filament profiles, we can correct for them. Assuming a Gaussian deconvolution (Eq. \ref{eqn:deconv}) does not accurately recover the intrinsic FWHM (Appendix \ref{sec:appendixTau}). Beam effects should be mitigated by forward modeling \citep[e.g.,][]{Juvela2012,Smith2014,Federrath2016,Seifried2017} or analytical functions that explicitly take these into account \citep{Fischera2012a}. Tools that go beyond the 1D description of a filament profile \citep[e.g., wavelet decomposition][]{Robitaille2014,Ossenkopf-Okada2019,Robitaille-2019} can allow for a deeper, more nuanced understanding of the nature of interstellar filaments and their environment \citep[e.g., as a hierarchical medium that is only limited by the beam size,][]{Robitaille2020}.

\begin{acknowledgements}
We thank K. Tassis and R. Skalidis for providing helpful comments. GVP acknowledges support by NASA through the NASA Hubble Fellowship grant  \#HST-HF2-51444.001-A  awarded  by  the  Space Telescope Science  Institute,  which  is  operated  by  the Association of Universities for Research in Astronomy, Incorporated, under NASA contract NAS5-26555. SEC acknowledges support by the National Science Foundation under Grant No. 2106607. 
This project has received funding from the European Research Council (ERC) under the European Union’s Horizon 2020 research and innovation programme (Grant agreement No. 851435). RJS gratefully acknowledges an STFC Ernest Rutherford fellowship (grant ST/N00485X/1). DS acknowledges support of the Bonn-Cologne Graduate School, which is funded through the German Excellence Initiative as well as funding by the Deutsche Forschungsgemeinschaft (DFG) via the Collaborative Research Center SFB 956 ``Conditions and Impact of Star Formation'' (subproject C6). This research has made use of data from the Herschel Gould Belt survey (HGBS) project (\url{http://gouldbelt-herschel.cea.fr}). The HGBS is a Herschel Key Programme jointly carried out by SPIRE Specialist Astronomy Group 3 (SAG 3), scientists of several institutes in the PACS Consortium (CEA Saclay, INAF-IFSI Rome and INAF-Arcetri, KU Leuven, MPIA Heidelberg), and scientists of the Herschel Science Center (HSC).
\end{acknowledgements}

\bibliographystyle{aa}
\bibliography{main}

\begin{thebibliography}{52}
\expandafter\ifx\csname natexlab\endcsname\relax\def\natexlab#1{#1}\fi

\bibitem[{{Alves} \& {Franco}(2007)}]{Alves2007}
{Alves}, F.~O. \& {Franco}, G.~A.~P. 2007, \aap, 470, 597

\bibitem[{{Andr{\'e}} {et~al.}(2014){Andr{\'e}}, {Di Francesco},
  {Ward-Thompson}, {Inutsuka}, {Pudritz}, \& {Pineda}}]{Andre2014}
{Andr{\'e}}, P., {Di Francesco}, J., {Ward-Thompson}, D., {et~al.} 2014, in
  Protostars and Planets VI, ed. H.~{Beuther}, R.~S. {Klessen}, C.~P.
  {Dullemond}, \& T.~{Henning}, 27

\bibitem[{{Arzoumanian} {et~al.}(2011){Arzoumanian}, {Andr{\'e}}, {Didelon},
  {K{\"o}nyves}, {Schneider}, {Men'shchikov}, {Sousbie}, {Zavagno}, {Bontemps},
  {di Francesco}, {Griffin}, {Hennemann}, {Hill}, {Kirk}, {Martin}, {Minier},
  {Molinari}, {Motte}, {Peretto}, {Pezzuto}, {Spinoglio}, {Ward-Thompson},
  {White}, \& {Wilson}}]{Arzoumanian2011}
{Arzoumanian}, D., {Andr{\'e}}, P., {Didelon}, P., {et~al.} 2011, \aap, 529, L6

\bibitem[{{Arzoumanian} {et~al.}(2019){Arzoumanian}, {Andr{\'e}},
  {K{\"o}nyves}, {Palmeirim}, {Roy}, {Schneider}, {Benedettini}, {Didelon}, {Di
  Francesco}, {Kirk}, \& {Ladjelate}}]{Arzoumanian2019}
{Arzoumanian}, D., {Andr{\'e}}, P., {K{\"o}nyves}, V., {et~al.} 2019, \aap,
  621, A42

\bibitem[{{Auddy} {et~al.}(2016){Auddy}, {Basu}, \& {Kudoh}}]{Auddy2016}
{Auddy}, S., {Basu}, S., \& {Kudoh}, T. 2016, \apj, 831, 46

\bibitem[{{Bensch} {et~al.}(2003){Bensch}, {Leuenhagen}, {Stutzki}, \&
  {Schieder}}]{Bensch2003}
{Bensch}, F., {Leuenhagen}, U., {Stutzki}, J., \& {Schieder}, R. 2003, \apj,
  591, 1013

\bibitem[{{Federrath}(2016)}]{Federrath2016}
{Federrath}, C. 2016, \mnras, 457, 375

\bibitem[{{Federrath} {et~al.}(2021){Federrath}, {Klessen}, {Iapichino}, \&
  {Beattie}}]{Federrath2021}
{Federrath}, C., {Klessen}, R.~S., {Iapichino}, L., \& {Beattie}, J.~R. 2021,
  Nature Astronomy, 5, 365

\bibitem[{{Fern{\'a}ndez-L{\'o}pez} {et~al.}(2014){Fern{\'a}ndez-L{\'o}pez},
  {Arce}, {Looney}, {Mundy}, {Storm}, {Teuben}, {Lee}, {Segura-Cox}, {Isella},
  {Tobin}, {Rosolowsky}, {Plunkett}, {Kwon}, {Kauffmann}, {Ostriker}, {Tassis},
  {Shirley}, \& {Pound}}]{Fernandez-Lopez2014}
{Fern{\'a}ndez-L{\'o}pez}, M., {Arce}, H.~G., {Looney}, L., {et~al.} 2014,
  \apjl, 790, L19

\bibitem[{{Fischera} \& {Martin}(2012{\natexlab{a}})}]{Fischera2012a}
{Fischera}, J. \& {Martin}, P.~G. 2012{\natexlab{a}}, \aap, 547, A86

\bibitem[{{Fischera} \& {Martin}(2012{\natexlab{b}})}]{Fischera2012b}
{Fischera}, J. \& {Martin}, P.~G. 2012{\natexlab{b}}, \aap, 542, A77

\bibitem[{{Franco}(1991)}]{Franco1991}
{Franco}, G.~A.~P. 1991, \aap, 251, 581

\bibitem[{{Gibb}(2008)}]{gibb2008}
{Gibb}, A.~G. 2008, {Star Formation in NGC 2068, NGC 2071, and Northern L1630},
  ed. B.~{Reipurth}, Vol.~4, 693

\bibitem[{Hacar {et~al.}(2018)Hacar, Tafalla, Forbrich, Alves, Meingast,
  Grossschedl, \& Teixeira}]{Hacar2018}
Hacar, A., Tafalla, M., Forbrich, J., {et~al.} 2018, \aap, 610, A77

\bibitem[{{Hacar} {et~al.}(2013){Hacar}, {Tafalla}, {Kauffmann}, \&
  {Kov{\'a}cs}}]{Hacar2013}
{Hacar}, A., {Tafalla}, M., {Kauffmann}, J., \& {Kov{\'a}cs}, A. 2013, \aap,
  554, A55

\bibitem[{{Harvey} {et~al.}(2008){Harvey}, {Huard}, {J{\o}rgensen},
  {Gutermuth}, {Mamajek}, {Bourke}, {Mer{\'\i}n}, {Cieza}, {Brooke}, {Chapman},
  {Alcal{\'a}}, {Allen}, {Evans}, {Di Francesco}, \& {Kirk}}]{Harvey2008}
{Harvey}, P.~M., {Huard}, T.~L., {J{\o}rgensen}, J.~K., {et~al.} 2008, \apj,
  680, 495

\bibitem[{{Hennebelle}(2013)}]{Hennebelle2013}
{Hennebelle}, P. 2013, \aap, 556, A153

\bibitem[{{Hennebelle} \& {Andr{\'e}}(2013)}]{HennebelleAndre2013}
{Hennebelle}, P. \& {Andr{\'e}}, P. 2013, \aap, 560, A68

\bibitem[{{Howard} {et~al.}(2019){Howard}, {Whitworth}, {Marsh}, {Clarke},
  {Griffin}, {Smith}, \& {Lomax}}]{Howard2019}
{Howard}, A.~D.~P., {Whitworth}, A.~P., {Marsh}, K.~A., {et~al.} 2019, \mnras,
  489, 962

\bibitem[{{Juvela} {et~al.}(2012{\natexlab{a}}){Juvela}, {Malinen}, \&
  {Lunttila}}]{Juvela2012}
{Juvela}, M., {Malinen}, J., \& {Lunttila}, T. 2012{\natexlab{a}}, \aap, 544,
  A141

\bibitem[{{Juvela} {et~al.}(2012{\natexlab{b}}){Juvela}, {Ristorcelli},
  {Pagani}, {Doi}, {Pelkonen}, {Marshall}, {Bernard}, {Falgarone}, {Malinen},
  {Marton}, {McGehee}, {Montier}, {Motte}, {Paladini}, {T{\'o}th}, {Ysard},
  {Zahorecz}, \& {Zavagno}}]{Juvela2012coldcores}
{Juvela}, M., {Ristorcelli}, I., {Pagani}, L., {et~al.} 2012{\natexlab{b}},
  \aap, 541, A12

\bibitem[{{Kenyon} {et~al.}(1994){Kenyon}, {Dobrzycka}, \&
  {Hartmann}}]{Kenyon1994}
{Kenyon}, S.~J., {Dobrzycka}, D., \& {Hartmann}, L. 1994, \aj, 108, 1872

\bibitem[{{Knude} \& {Hog}(1998)}]{Knude1998}
{Knude}, J. \& {Hog}, E. 1998, \aap, 338, 897

\bibitem[{{K{\"o}nyves} {et~al.}(2015){K{\"o}nyves}, {Andr{\'e}},
  {Men'shchikov}, {Palmeirim}, {Arzoumanian}, {Schneider}, {Roy}, {Didelon},
  {Maury}, {Shimajiri}, {Di Francesco}, {Bontemps}, {Peretto}, {Benedettini},
  {Bernard}, {Elia}, {Griffin}, {Hill}, {Kirk}, {Ladjelate}, {Marsh}, {Martin},
  {Motte}, {Nguy{\^e}n Luong}, {Pezzuto}, {Roussel}, {Rygl}, {Sadavoy},
  {Schisano}, {Spinoglio}, {Ward-Thompson}, \& {White}}]{Konyves2015}
{K{\"o}nyves}, V., {Andr{\'e}}, P., {Men'shchikov}, A., {et~al.} 2015, \aap,
  584, A91

\bibitem[{{Lada} {et~al.}(1999){Lada}, {Alves}, \& {Lada}}]{Lada1999}
{Lada}, C.~J., {Alves}, J., \& {Lada}, E.~A. 1999, \apj, 512, 250

\bibitem[{{Leike} {et~al.}(2020){Leike}, {Glatzle}, \&
  {En{\ss}lin}}]{Leike2020}
{Leike}, R.~H., {Glatzle}, M., \& {En{\ss}lin}, T.~A. 2020, \aap, 639, A138

\bibitem[{{Louvet} {et~al.}(2021){Louvet}, {Hennebelle}, {Men'shchikov},
  {Didelon}, {Ntormousi}, \& {Motte}}]{Louvet2021}
{Louvet}, F., {Hennebelle}, P., {Men'shchikov}, A., {et~al.} 2021, \aap, 653,
  A157

\bibitem[{{Monsch} {et~al.}(2018){Monsch}, {Pineda}, {Liu}, {Zucker}, {How-Huan
  Chen}, {Pattle}, {Offner}, {Di Francesco}, {Ginsburg}, {Ercolano}, {Arce},
  {Friesen}, {Kirk}, {Caselli}, \& {Goodman}}]{Monsch2018}
{Monsch}, K., {Pineda}, J.~E., {Liu}, H.~B., {et~al.} 2018, \apj, 861, 77

\bibitem[{{Ortiz-Le{\'o}n} {et~al.}(2018){Ortiz-Le{\'o}n}, {Loinard}, {Dzib},
  {Kounkel}, {Galli}, {Tobin}, {Evans}, {Hartmann}, {Rodr{\'\i}guez},
  {Brice{\~n}o}, {Torres}, \& {Mioduszewski}}]{Ortiz-Leon2018}
{Ortiz-Le{\'o}n}, G.~N., {Loinard}, L., {Dzib}, S.~A., {et~al.} 2018, \apjl,
  869, L33

\bibitem[{{Ossenkopf-Okada} \& {Stepanov}(2019)}]{Ossenkopf-Okada2019}
{Ossenkopf-Okada}, V. \& {Stepanov}, R. 2019, \aap, 621, A5

\bibitem[{{Ostriker}(1964)}]{Ostriker1964}
{Ostriker}, J. 1964, \apj, 140, 1056

\bibitem[{{Panopoulou} {et~al.}(2021){Panopoulou}, {Dickinson}, {Readhead},
  {Pearson}, \& {Peel}}]{panopoulou2021}
{Panopoulou}, G.~V., {Dickinson}, C., {Readhead}, A.~C.~S., {Pearson}, T.~J.,
  \& {Peel}, M.~W. 2021, ApJ, in press, arXiv:2106.14267

\bibitem[{{Panopoulou} {et~al.}(2017){Panopoulou}, {Psaradaki}, {Skalidis},
  {Tassis}, \& {Andrews}}]{Panopoulou2017}
{Panopoulou}, G.~V., {Psaradaki}, I., {Skalidis}, R., {Tassis}, K., \&
  {Andrews}, J.~J. 2017, \mnras, 466, 2529

\bibitem[{{Panopoulou} {et~al.}(2016){Panopoulou}, {Psaradaki}, \&
  {Tassis}}]{Panopoulou2016a}
{Panopoulou}, G.~V., {Psaradaki}, I., \& {Tassis}, K. 2016, \mnras, 462, 1517

\bibitem[{{Priestley} \& {Whitworth}(2021)}]{Priestley2021}
{Priestley}, F.~D. \& {Whitworth}, A.~P. 2021, \mnras
  [\eprint[arXiv]{2109.13277}]

\bibitem[{{Rivera-Ingraham} {et~al.}(2016){Rivera-Ingraham}, {Ristorcelli},
  {Juvela}, {Montillaud}, {Men'shchikov}, {Malinen}, {Pelkonen}, {Marston},
  {Martin}, {Pagani}, {Paladini}, {Paradis}, {Ysard}, {Ward-Thompson},
  {Bernard}, {Marshall}, {Montier}, \& {T{\'o}th}}]{RiveraIngraham2016}
{Rivera-Ingraham}, A., {Ristorcelli}, I., {Juvela}, M., {et~al.} 2016, \aap,
  591, A90

\bibitem[{{Robitaille} {et~al.}(2020){Robitaille}, {Abdeldayem}, {Joncour},
  {Moraux}, {Motte}, {Lesaffre}, \& {Khalil}}]{Robitaille2020}
{Robitaille}, J.~F., {Abdeldayem}, A., {Joncour}, I., {et~al.} 2020, \aap, 641,
  A138

\bibitem[{{Robitaille} {et~al.}(2014){Robitaille}, {Joncas}, \&
  {Miville-Desch{\^e}nes}}]{Robitaille2014}
{Robitaille}, J.~F., {Joncas}, G., \& {Miville-Desch{\^e}nes}, M.~A. 2014,
  \mnras, 440, 2726

\bibitem[{{Robitaille} {et~al.}(2019){Robitaille}, {Motte}, {Schneider},
  {Elia}, \& {Bontemps}}]{Robitaille-2019}
{Robitaille}, J.~F., {Motte}, F., {Schneider}, N., {Elia}, D., \& {Bontemps},
  S. 2019, \aap, 628, A33

\bibitem[{{Schisano} {et~al.}(2014){Schisano}, {Rygl}, {Molinari}, {Busquet},
  {Elia}, {Pestalozzi}, {Polychroni}, {Billot}, {Carey}, {Paladini},
  {Noriega-Crespo}, {Moore}, {Plume}, {Glover}, \&
  {V{\'a}zquez-Semadeni}}]{Schisano2014}
{Schisano}, E., {Rygl}, K.~L.~J., {Molinari}, S., {et~al.} 2014, \apj, 791, 27

\bibitem[{{Schlafly} {et~al.}(2014){Schlafly}, {Green}, {Finkbeiner}, {Rix},
  {Bell}, {Burgett}, {Chambers}, {Draper}, {Hodapp}, {Kaiser}, {Magnier},
  {Martin}, {Metcalfe}, {Price}, \& {Tonry}}]{schlafly2014}
{Schlafly}, E.~F., {Green}, G., {Finkbeiner}, D.~P., {et~al.} 2014, \apj, 786,
  29

\bibitem[{{Seifried} {et~al.}(2017){Seifried}, {S{\'a}nchez-Monge}, {Suri}, \&
  {Walch}}]{Seifried2017}
{Seifried}, D., {S{\'a}nchez-Monge}, {\'A}., {Suri}, S., \& {Walch}, S. 2017,
  \mnras, 467, 4467

\bibitem[{{Smith} {et~al.}(2014){Smith}, {Glover}, \& {Klessen}}]{Smith2014}
{Smith}, R.~J., {Glover}, S. C.~O., \& {Klessen}, R.~S. 2014, \mnras, 445, 2900

\bibitem[{{Sousbie}(2011)}]{Sousbie2011}
{Sousbie}, T. 2011, \mnras, 414, 350

\bibitem[{{Strai{\v{z}}ys} {et~al.}(1996){Strai{\v{z}}ys}, {{\v{C}}ernis}, \&
  {Barta{\v{s}}i{\={u}}t{\.{e}}}}]{straizys1996}
{Strai{\v{z}}ys}, V., {{\v{C}}ernis}, K., \& {Barta{\v{s}}i{\={u}}t{\.{e}}}, S.
  1996, Baltic Astronomy, 5, 125

\bibitem[{{Suri} {et~al.}(2019){Suri}, {S{\'a}nchez-Monge}, {Schilke},
  {Clarke}, {Smith}, {Ossenkopf-Okada}, {Klessen}, {Padoan}, {Goldsmith},
  {Arce}, {Bally}, {Carpenter}, {Ginsburg}, {Johnstone}, {Kauffmann}, {Kong},
  {Lis}, {Mairs}, {Pillai}, {Pineda}, \& {Duarte-Cabral}}]{Suri2019}
{Suri}, S., {S{\'a}nchez-Monge}, {\'A}., {Schilke}, P., {et~al.} 2019, \aap,
  623, A142

\bibitem[{{Verschuur}(1993)}]{Verschuur93}
{Verschuur}, G.~L. 1993, \aj, 106, 2580

\bibitem[{{Ward-Thompson} {et~al.}(2010){Ward-Thompson}, {Kirk}, {Andr{\'e}},
  {Saraceno}, {Didelon}, {K{\"o}nyves}, {Schneider}, {Abergel}, {Baluteau},
  {Bernard}, {Bontemps}, {Cambr{\'e}sy}, {Cox}, {di Francesco}, {di Giorgio},
  {Griffin}, {Hargrave}, {Huang}, {Li}, {Martin}, {Men'shchikov}, {Minier},
  {Molinari}, {Motte}, {Olofsson}, {Pezzuto}, {Russeil}, {Sauvage},
  {Sibthorpe}, {Spinoglio}, {Testi}, {White}, {Wilson}, {Woodcraft}, \&
  {Zavagno}}]{ward-thompson2010}
{Ward-Thompson}, D., {Kirk}, J.~M., {Andr{\'e}}, P., {et~al.} 2010, \aap, 518,
  L92

\bibitem[{{Whitworth} {et~al.}(2021){Whitworth}, {Priestley}, \&
  {Arzoumanian}}]{Whitowrth2021}
{Whitworth}, A.~P., {Priestley}, F.~D., \& {Arzoumanian}, D. 2021, \mnras, 508,
  2736

\bibitem[{{Zagury} {et~al.}(1999){Zagury}, {Boulanger}, \&
  {Banchet}}]{Zagury1999}
{Zagury}, F., {Boulanger}, F., \& {Banchet}, V. 1999, \aap, 352, 645

\bibitem[{{Zucker} \& {Chen}(2018)}]{Zucker2018Radfil}
{Zucker}, C. \& {Chen}, H. H.-H. 2018, \apj, 864, 152

\bibitem[{{Zucker} {et~al.}(2020){Zucker}, {Speagle}, {Schlafly}, {Green},
  {Finkbeiner}, {Goodman}, \& {Alves}}]{Zucker2020}
{Zucker}, C., {Speagle}, J.~S., {Schlafly}, E.~F., {et~al.} 2020, \aap, 633,
  A51

\end{thebibliography}

\begin{appendix}

\section{Cloud distances}
\label{sec:appendix}

\begin{figure*}
    \centering
    \includegraphics[scale=1]{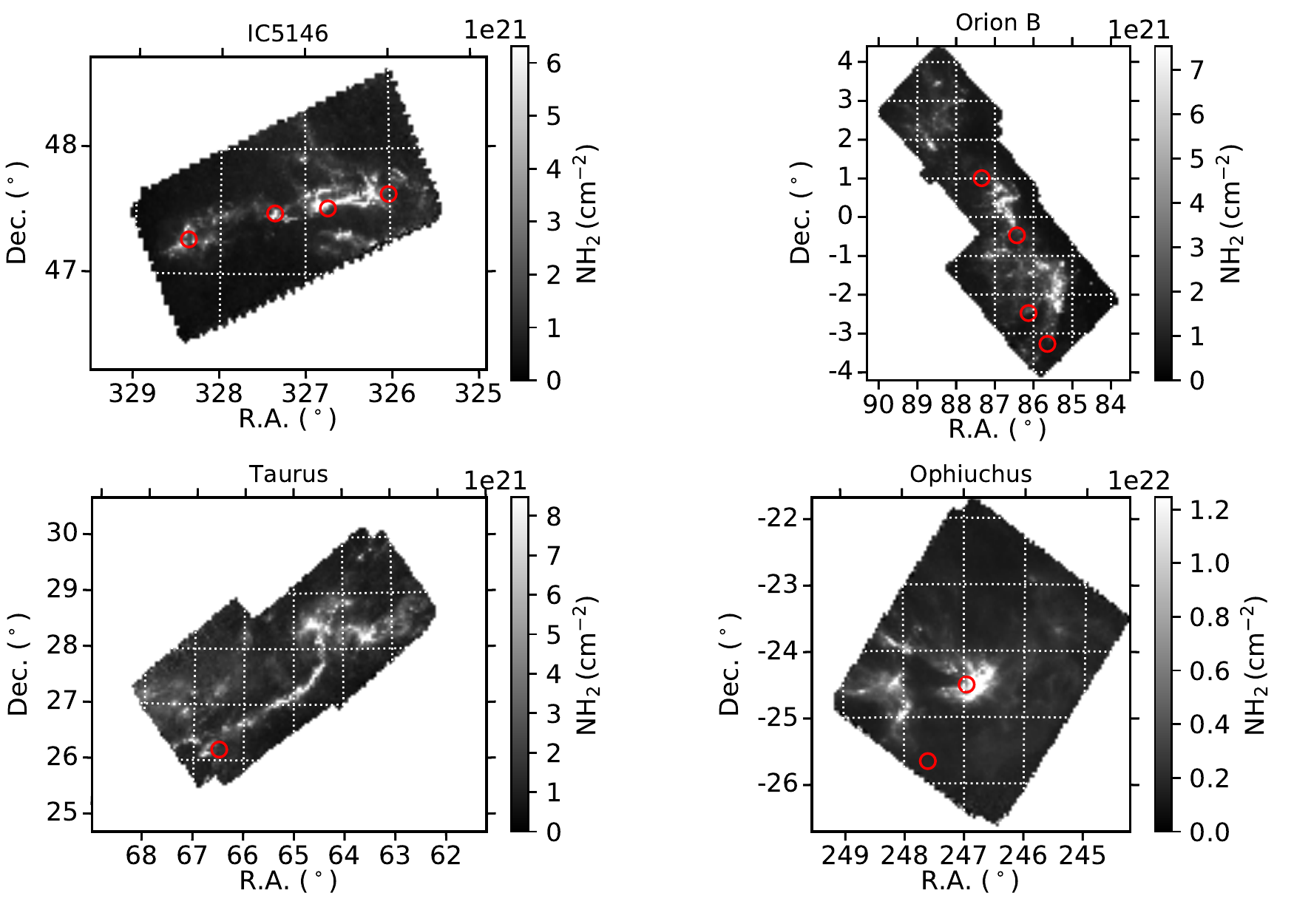}
    \caption{\textit{Herschel} column density maps of clouds with distance measurements from \citet{Zucker2020}. Red circles mark the sightlines with measured distance.}
    \label{fig:maps}
\end{figure*}

We used the latest results from 3D dust extinction mapping to update the distance estimates of clouds in the sample of \citet{Arzoumanian2019}. \citet{Zucker2020} provide distance measurements for a large list of nearby molecular clouds. They combined stellar photometry with the \textit{Gaia} Data-Release-2 stellar parallax to model stellar extinction as a function of distance in discrete sightlines toward these clouds. For each cloud we considered all sightlines (from their table A.1) that fall within the footprint of the \textit{Herschel} maps (Fig. \ref{fig:maps}). We combined the systematic and statistical uncertainties for each measurement and compared the $\pm 1\sigma$ ranges among all sightlines. We considered the lower limit of the cloud distance as the minimum of these bounds, and similarly for the upper limit. We thus obtained a lower and upper limit for the distance to four clouds in the sample: IC 5146, Orion B, Taurus, and Ophiuchus.

\begin{figure*}
    \centering
    \includegraphics[scale=1]{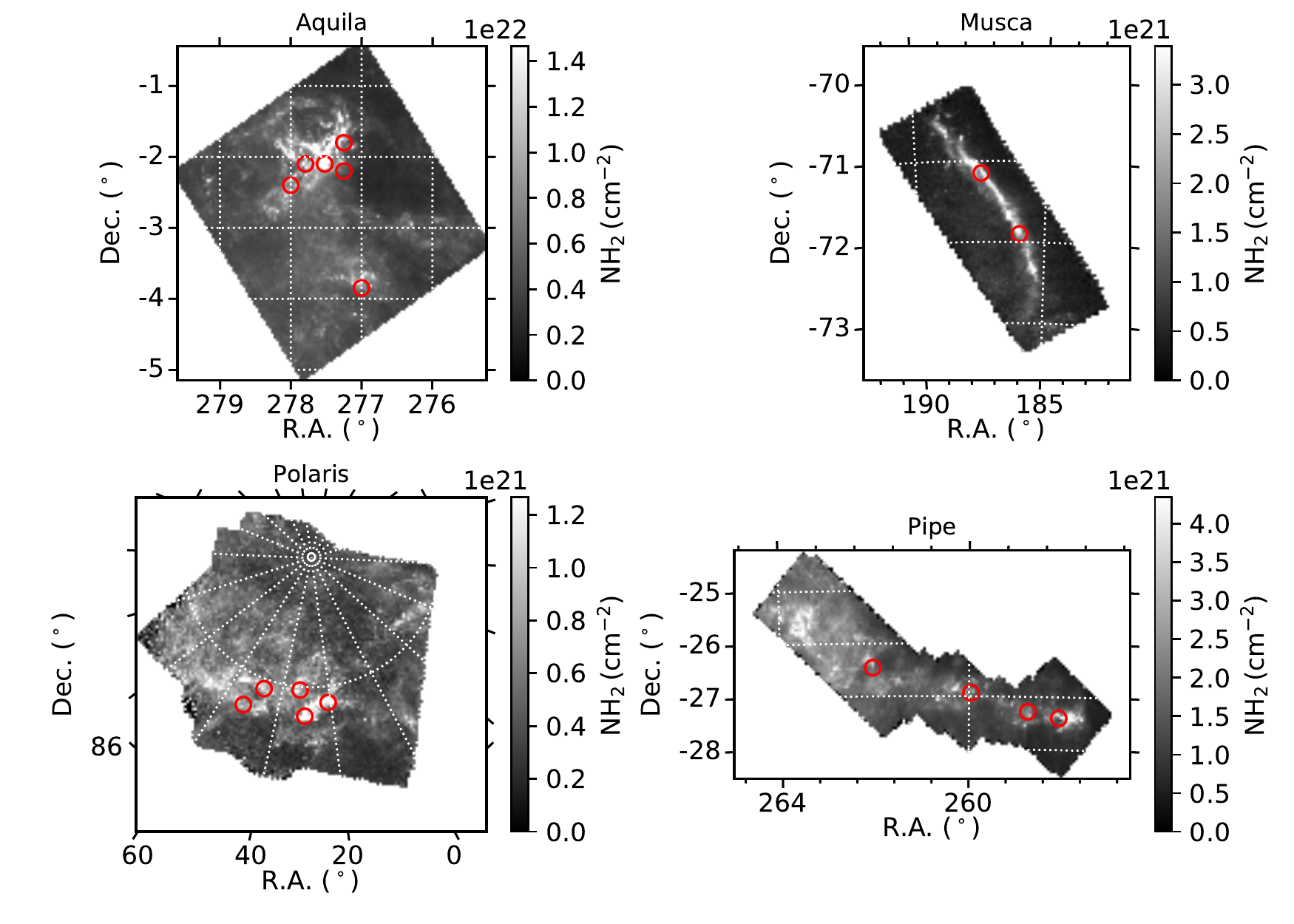}
    \caption{\textit{Herschel} column density maps of clouds for which we found the distance using the 3D dust extinction map of \citet{Leike2020}. Red circles mark the sightlines used to derive distances.}
    \label{fig:maps_new}
\end{figure*}

For the remaining four clouds in the sample (Aquila, Musca, Polaris, and Pipe), there are no suitable distance estimates from \citet{Zucker2020}, that is to say either no sightline passes within the \textit{Herschel} footprint, or the cloud was not studied at all, as for Musca. We therefore analyzed 3D dust maps to calculate distance limits.
We used the \citet{Leike2020} map which offers the best distance resolution among available maps ($\sim$ 1 pc). First, we selected lines of sight that overlap with filaments as identified in \citet{Arzoumanian2019} (shown in Fig. \ref{fig:maps_new}). Using the \texttt{dustmaps} python package, we queried the \citet{Leike2020} map to obtain the differential optical depth (optical depth per parsec) for each line of sight. We converted the differential optical depth to differential $G-$band extinction, $\delta{A_G}$, following \citet{panopoulou2021}. For all sightlines, the differential optical depth shows a prominent peak at a certain distance (Fig. \ref{fig:distances}). We found the distance of the $\delta{A_G}$ peak for each sightline. The minimum and maximum peak locations for each cloud are considered to be the lower and upper limits for its distance.

Distance measurements for the sightlines toward all clouds in the sample are given in Table \ref{tab:clouddist}. For measurements from \citet{Zucker2020}, the quoted distance uncertainties include the full statistical and systematic uncertainty as provided in their table A.1. For measurements from this work, we quote the uncertainty determined by the chosen distance binning of the 3D dust map (5 pc). In the following, we compare the updated distance estimates with literature values used in \citet{Arzoumanian2019}.

\subsection*{IC5146}
The distance range for this cloud is [686, 833] pc by combining the results of \citet{Zucker2020}, as described above. 
The original distance adopted by \citet{Arzoumanian2011} of 460$^{+40}_{-60}$~pc was derived by \citet{Lada1999} using star counts. An alternative distance of 950$\pm$ 80~pc was also considered \citep[appendix A of][]{Arzoumanian2011}, which was derived by stellar photometry of late B-type members of the IC 5146 cluster \citep{Harvey2008}. 

\subsection*{Orion B}
The distance range for this cloud is [386, 474] pc \citep[from][]{Zucker2020}. This is consistent with the distance of 400 pc adopted in \citet{Arzoumanian2019} \citep[obtained from][]{gibb2008}. 

\subsection*{Polaris Flare}
\begin{figure}
    \centering
    \includegraphics[scale=1]{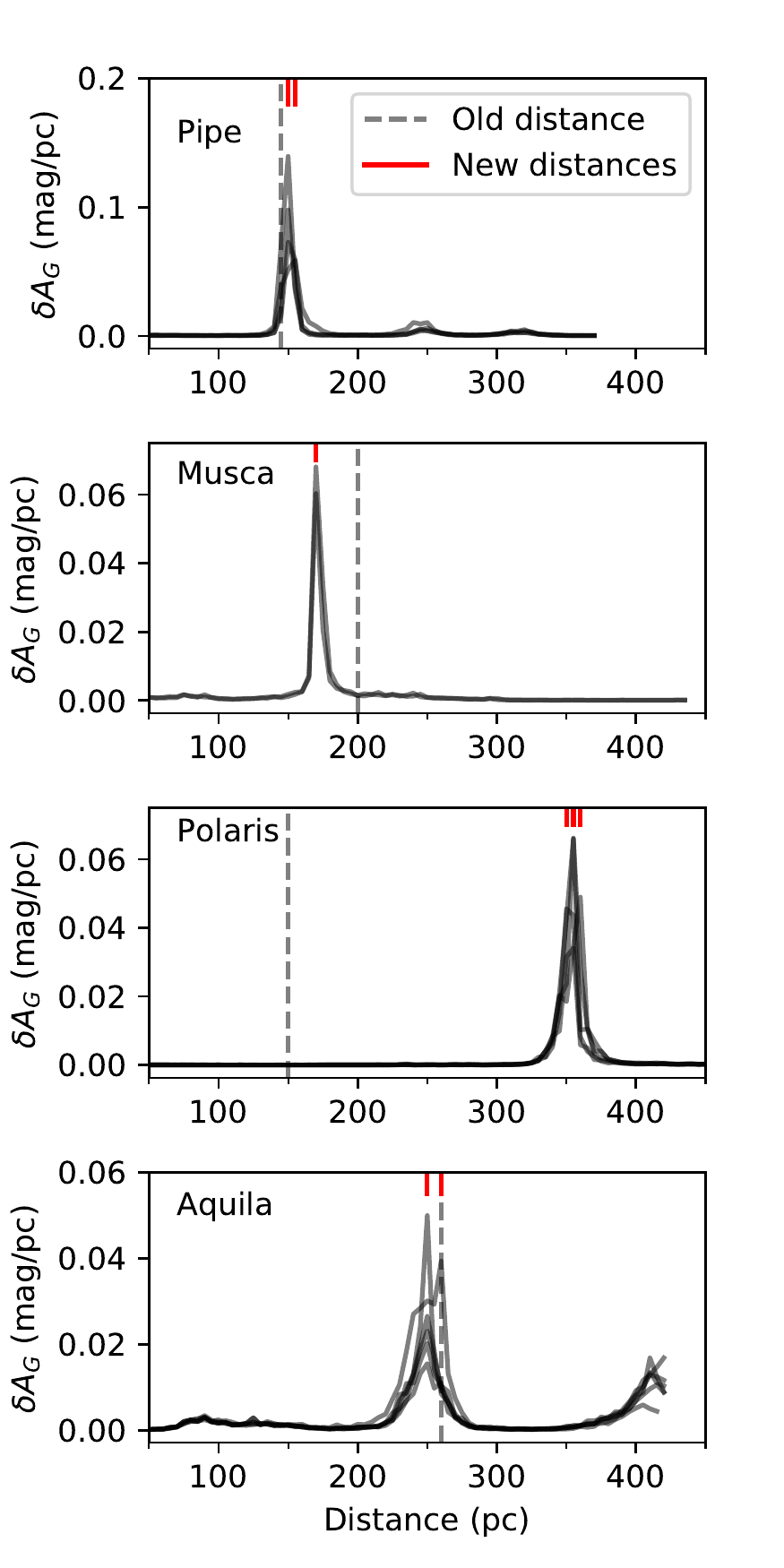}
    \caption{Distance determination for the Pipe Nebula, Musca, Polaris Flare, and Aquila Rift clouds. Each subpanel shows the differential $G-band$ extinction per parsec as a function of distance for different sightlines within the \textit{Herschel} map of each cloud (transparent black lines). Old distance estimates \citep[adopted in][]{Arzoumanian2019} are shown as a vertical dashed line, while the new distance measurement for each line of sight is shown as a red vertical segment at the top of each panel. }
    \label{fig:distances}
\end{figure}

Our distance estimate for the Polaris Flare is [350, 360] pc. This is significantly discrepant from the distance of 150 pc assumed in \citet{Arzoumanian2019}. Despite this being a commonly adopted distance \citep[e.g.,][]{Bensch2003,ward-thompson2010}, it is nevertheless incorrect, and can be traced back to \citet{Zagury1999}, who assumed that the cloud is foreground to Polaris (the star) \citep[see][for a detailed literature review]{schlafly2014}. Polaris (the star) exhibits some extinction and polarization, leading \citet{Panopoulou2016a} to also erroneously conclude that the cloud is in front of the North star. However the bulk of extinction clearly arises at 350 pc as seen in Fig. \ref{fig:distances} \citep[and also][]{schlafly2014}. 

\subsection*{Pipe}
Extinction toward the Pipe nebula is found to lie within [150, 155] pc. This is consistent with the previously adopted distance of 145$\pm$16 pc, based on linear polarimetry and Hipparcos distances \citep{Alves2007}.

\subsection*{Aquila}
Our distance limits to Aquila are [250, 260] pc, consistent with the distance of 260$\pm 37$  pc, which was based on stellar extinctions in the general area of Serpens \citep{straizys1996}.

\subsection*{Musca}
Our determination of the distance to Musca is 170 pc; there is no variation among the selected sightlines. The previously adopted distance was 200 pc, selected by Cox et al to be "in between" the estimates by \citet{Franco1991} and \citet{Knude1998} (stellar extinctions toward the general region of the Chamaeleon clouds).

\subsection*{Taurus L1495}
The distance limits from \citet{Zucker2020} are [120, 140] pc, which is consistent with the previously adopted distance of 140 pc \citep[e.g., based on stellar photometry from][]{Kenyon1994}.

\subsection*{Ophiuchus}
The distance limits for Ophiuchus of [131, 145] pc are consistent with the previously adopted distance of 140 pc \citep[in turn based on highly accurate distance measurements toward parts of the cloud of 138$\pm 3$ pc and 144$\pm 1$ pc from Very Long Baseline Interferometry and Gaia,][]{Ortiz-Leon2018}.

\begin{table}
    \centering
    \begin{tabular}{c|c|c|c|c|c}
         Cloud & l ($^\circ$)& b ($^\circ$) & $d_{\rm new}$ (pc) & $d_{\rm old}$ (pc)& Ref.\\
         \hline
         IC 5146 & 93.7 & -4.6 & 774$^{+40}_{-41}$ & 460 & (1)\\
         IC 5146 & 93.4 & -4.2 & 792$^{+41}_{-42}$ & 460 & (1)\\
         IC 5146 & 94.0 & -4.9 & 730$^{+44}_{-41}$ & 460 & (1)\\
         IC 5146 & 94.4 & -5.5 & 751$^{+38}_{-39}$ & 460 & (1)\\
         Orion B & 205.7 & -14.8 & 436$^{+32}_{-31}$ & 460 & (1)\\
         Orion B & 207.9 & -16.8 & 411$^{+22}_{-24}$ & 460 & (1)\\
         Orion B & 207.4 & -16.0 & 451$^{+23}_{-22}$ & 460 & (1)\\
         Orion B & 204.8 & -13.3 & 415$^{+20}_{-21}$ & 460 & (1)\\
         Taurus  & 171.6 & -15.8 & 130$^{+11}_{-10}$ & 140 &(1)\\
         Ophiuchus & 352.7 & 15.4 & 139$^{+7}_{-6}$ & 140 &(1)\\
         Ophiuchus & 353.2 & 16.6 & 139$^{+7}_{-7}$ & 140 &(1)\\
         Aquila & 28.4 & 3.9 & 250$\pm 5$ & 260 &(2)\\
         Aquila & 28.8 & 4.1 & 250$\pm 5$ & 260 &(2)\\
         Aquila & 28.6 & 3.2 & 250$\pm 5$ & 260 &(2)\\
         Aquila & 28.6 & 3.8 & 250$\pm 5$ & 260 &(2)\\
         Aquila & 28.7 & 3.5 & 250$\pm 5$ & 260 &(2)\\
         Aquila & 26.8 & 3.4 & 260$\pm 5$ & 260 &(2)\\
         Polaris & 124.3 & 25.3 & 355$\pm 5$ & 150 &(2)\\
         Polaris & 123.7 & 24.8 & 355$\pm 5$ & 150 &(2)\\
         Polaris & 123.7 & 25.2 & 355$\pm 5$ & 150 &(2)\\
         Polaris & 123.3 & 24.9 & 350$\pm 5$ & 150 &(2)\\
         Polaris & 124.7 & 25.2 & 360$\pm 5$ & 150 &(2)\\
         Musca & 301.2 & -8.4 & 170$\pm 5$ & 200 &(2)\\
         Musca & 300.8 & -9.1 & 170$\pm 5$ & 200 &(2)\\
         Pipe & 0.0 & 4.6 & 155$\pm 5$ & 145 &(2)\\
         Pipe & 358.6 & 5.9 & 150$\pm 5$ & 145 &(2)\\
         Pipe & 357.2 & 6.9 & 150$\pm 5$ & 145 &(2)\\
         Pipe & 357.7 & 6.6 & 150$\pm 5$ & 145 &(2)\\
    \end{tabular}
    \caption{Summary of sightlines used to determine distance limits to clouds. Galactic coordinates are provided, as well as the distance estimates for each sightline. References for the new distance limits: (1) \citet{Zucker2020}, (2) this work. }
    \label{tab:clouddist}
\end{table}

\section{A case study in Taurus on the effects of beam convolution}
\label{sec:appendixTau}

We investigated the effect of resolution on a radial profile from the Taurus B211/B213 filament. First, we convolved the column density map of Taurus with Gaussian kernels of different sizes to simulate the effect of observing the same cloud at lower resolution. We chose resolutions in physical size that correspond to the \textit{Herschel} beam of 18.2\arcsec\, at distances of 140 pc (the native resolution), 260 pc (distance to Aquila), 423 pc (distance to Orion), and 762 pc (distance to IC 5146).

The convolution was performed using the \texttt{astropy} \texttt{convolve} method. A 2D Gaussian kernel was defined so that the final resolution of the image (beam size) has a physical size of 0.023 pc, 0.037 pc, and 0.067 pc, corresponding to the \textit{Herschel} beam of 18.2\arcsec observed at distances of 260 pc, 423 pc, and 762 pc. More specifically, we calculated the kernel standard deviation as: $\rm \sigma_{kernel} = \sqrt{\rm beam^2_{final} - beam^2_{initial}}/(2\sqrt{2\ln{2}})$, where $\rm beam_{final}$ is the desired resolution of the image, $\rm beam_{initial}$ is the \textit{Herschel} beam, and all quantities are measured in units of pixel on the image.

\begin{figure*}
    \centering
    \includegraphics[scale=0.85]{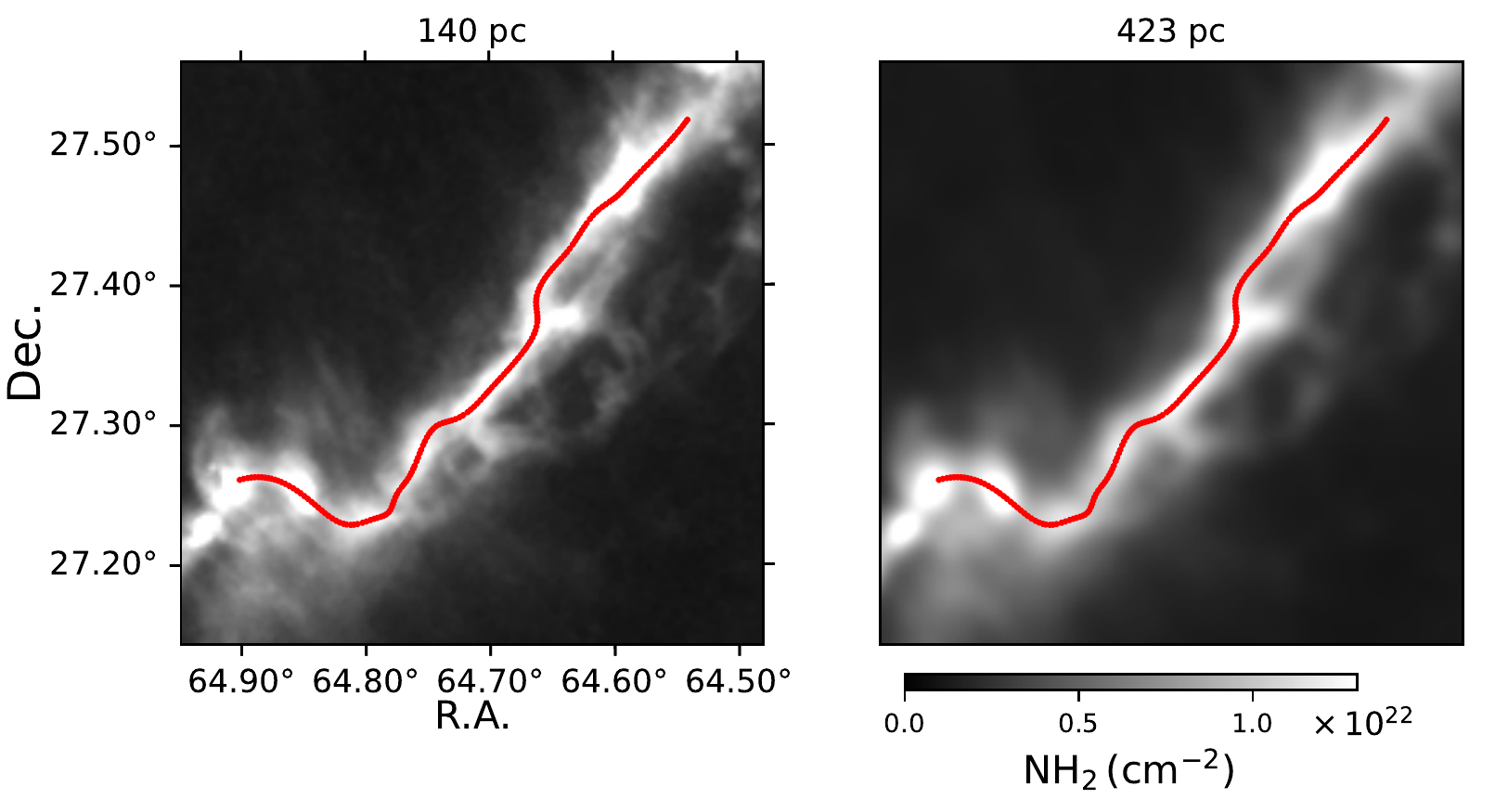}
    \caption{Filament in Taurus observed with  different beam sizes. The filament crest (obtained from the HGBS archive) is marked with a red line. The images were obtained by convolving the original \textit{Herschel} column density image of Taurus (left panel) with a Gaussian kernel to achieve effective resolutions corresponding distances of 260 - 762 pc. The right panel shows the map for 423 pc.}
    \label{fig:taurus_prof_map}
\end{figure*}

We used the publicly available skeleton of the Taurus \textit{Herschel} map, which was produced using \texttt{DisPerSe} -- see \citet{Arzoumanian2019}. We input this skeleton to the \texttt{radfil} Python package \citep{Zucker2018Radfil} and extracted the median radial profile along the filament crest (shown in Fig. \ref{fig:taurus_prof_map}). We focused on a single filament to isolate the effect of resolution from other effects such as averaging over the filament population.

\begin{figure*}
    \centering
    \includegraphics{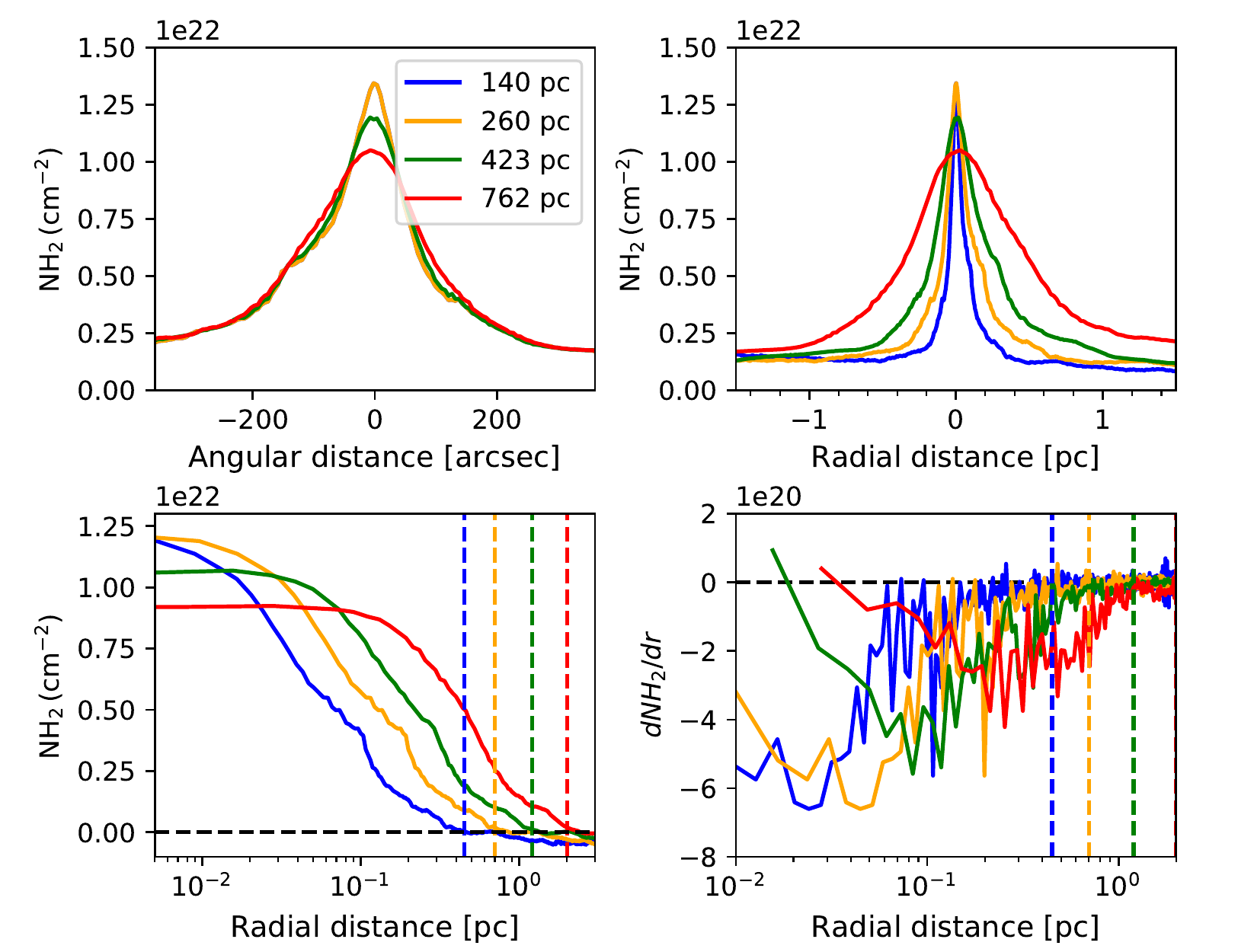}
    \caption{Median radial profile of a filament in Taurus (Fig. \ref{fig:taurus_prof_map}) for different choices of resolution, corresponding to the \textit{Herschel} beam of 18.2\arcsec \, at distances of  140 pc, blue; 260 pc, orange; 423 pc, green; and 762 pc, red. Top left: Profile comparison in angular units. Top right: Profile comparison in physical units. Bottom left: Profiles after background subtraction (with logarithmic horizontal axis for better visualization). The vertical dashed lines mark the $r_{out}$ defined in the bottom right panel. Bottom right: Derivative of radial profile. The radial distance where the derivative flattens determines $r_{out}$ (vertical dashed lines), following \citet{Arzoumanian2019}.}
    \label{fig:taurus}
\end{figure*}

The median radial profile obtained from the images "observed" with lower angular resolution is shown in Fig. \ref{fig:taurus}. By changing the resolution, we see a reduction in the peak amplitude of the profile, while the overall shape remains similar in {angular units} (Fig. \ref{fig:taurus}, left panel). When comparing the profiles in physical units, the profiles become drastically broader for larger beam sizes (lower resolution).

For each profile, we measured the FWHM similarly to \cite{Arzoumanian2019}, as follows. We found the outer truncation radius, $r_{out}$, where the derivative of the profile, $dNH_2/dr,$ becomes consistent with zero over a range of distances. The derivative and corresponding $r_{out}$ are shown in Fig. \ref{fig:taurus} (bottom right panel). The values of $r_{out}$ are 0.25 pc, 0.5 pc, 0.95 pc, and 1.5 pc for the profiles at 140 pc, 260 pc, 423 pc, and 762 pc. We then calculated the mean column density of the profile at all radial distances beyond $r_{out}$. We subtracted this mean value (background) from the profile and then set all negative values to zero. We finally found the half-radius and multiplied by 2 to obtain $\rm FWHM_{obs}$. The results are shown in Fig. \ref{fig:tau_dist} and discussed in the main text.

\begin{figure}
    \centering
    \includegraphics[scale=1]{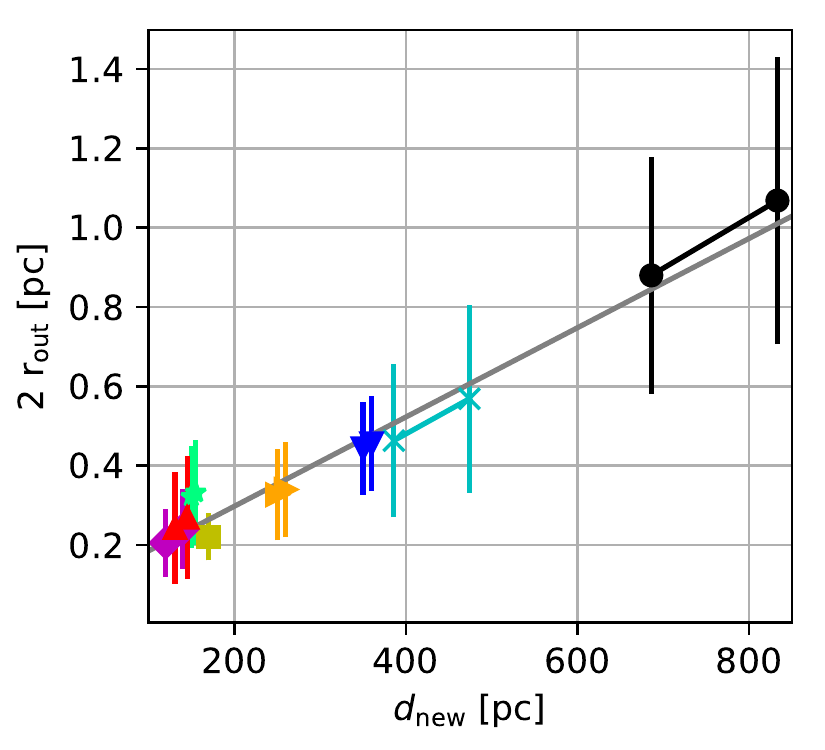}
    \caption{Outer truncation diameter as a function of (updated) cloud distance. The line is a linear fit to the data points choosing all lower distance limits. A fit to all upper distance limits yields similar results. Symbols as in Fig. \ref{fig:wa_dist}.}
    \label{fig:rout}
\end{figure}

Previous works have indicated that filament widths have a dependence on the choice of the truncation radius \citep{Smith2014, Panopoulou2017}. To ensure a fair comparison with \citet{Arzoumanian2019}, we compared our recovered $r_{out}$ with the reported per-cloud average $r_{out}$ from their work. We rescaled the values of $2\,\,r_{\rm out}$ \citep[][table 3]{Arzoumanian2019} to the new cloud distances and plotted them in Fig. \ref{fig:rout}. The values were found to scale approximately linearly with distance. We fit a linear regression to the data using the lower distance limits and obtained the following:
\begin{equation}
    2 \,\, r_{\rm out} = 0.00112 \,\, d_{\rm new} + 0.072 \, \rm pc.
    \label{eq:rout}
\end{equation}
 Fitting the lower or upper limits of $d_{\rm new}$ yields essentially identical results. Dividing by 2, we obtained the slope and intercept for obtaining the truncation radius $r_{\rm out}$ at a given distance. We checked that using the mean $r_{out}$ for a given distance from eq. \ref{eq:rout} produces essentially identical $\rm FWHM_{obs}$ with respect to those that used the profile-flattening criterion for the example filament in our resolution study. The one exception is for the largest distance considered, where the Taurus profile differs by 40\% for the two choices of $r_{out}$.

The effect of the beam size has previously been treated as a simple convolution of Gaussians (Eq. \ref{eqn:deconv}). However, Eq. \ref{eqn:deconv} should not be used to deconvolve any arbitrary functional form of a profile from a Gaussian beam. This can be readily understood as a consequence of the Fourier properties of a Gaussian function and the convolution theorem. Indeed, as noted in \citet{Zucker2018Radfil}, Eq. \ref{eqn:deconv} does not have the desired effect of correcting for beam convolution. For the example profiles shown in Fig. \ref{fig:taurus}, the "deconvolved" FWHM$_{\rm dec}$ from Eq. \ref{eqn:deconv} can differ by up to a factor of 10 from the initial FWHM$_{\rm obs}$ of the profile at 140 pc. A simple beam deconvolution does not recover intrinsic properties of the profile.

\end{appendix}
\end{document}